# A Processing Route to Chalcogenide Perovskites Alloys with Tunable Band Gap via Anion Exchange

Kevin Ye[1*], Ida Sadeghi[1*], Michael Xu[1], Jack Van Sambeek[1], Tao Cai[1], Jessica Dong[1], Rishabh Kothari,[1] James M. LeBeau[1], R. Jaramillo[1‡]

1. Department of Materials Science and Engineering, Massachusetts Institute of Technology, Cambridge, MA 02139, USA

*These authors contributed equally to this work

‡ *email address: rjaramil@mit.edu*

**Abstract**
We demonstrate synthesis of BaZr(S,Se)$_3$ chalcogenide perovskite alloys by selenization of BaZrS$_3$ thin films. The anion-exchange process produces films with tunable composition and band gap without changing the orthorhombic perovskite crystal structure or the film microstructure. The direct band gap is tunable between 1.5 and 1.9 eV. The alloy films made in this way feature 100 × stronger photoconductive response and a lower density of extended defects, compared to alloy films made by direct growth. The perovskite structure is stable in high-selenium-content thin films with and without epitaxy. The manufacturing-compatible process of selenization in H$_2$Se gas may spur the development of chalcogenide perovskite solar cell technology.

## 1. Introduction

Tuning semiconductor material properties by well-controlled alloying processes is enabling for device design and optimization. Manufacturing-compatible processes to reliably tune the band gap are particularly important for applications in optoelectronics and photovoltaics, and are well-developed for Si-Ge, III-V, and II-VI compounds. Compared to these semiconductors, chalcogenide perovskites are not well understood or well controlled. Nevertheless, they may be of interest for photovoltaics.[1–6] These materials are comprised of non-toxic, Earth-abundant elements, and are thermally stable in air at least up to 550 °C, which is attractive from the standpoint of future manufacturing and deployment.[7,8] The most well-studied is BaZrS$_3$ (orthorhombic corner-sharing structure type, space group *Pnma*, no. 62), which is a direct band gap semiconductor with band gap $E_g$ = 1.9 eV.[9] BaZrS$_3$ features strong dielectric response, strong optical absorption, band-edge photoluminescence, and relatively slow non-radiative electron-hole recombination.[3–5,10–12]

Theory predicts several routes to lowering the band gap of BaZrS$_3$ by alloying, and Se- and Ti-alloying have been demonstrated experimentally.[4,13–16] We recently reported synthesis of chalcogenide perovskite alloy epitaxial thin films spanning the full composition range BaZrS$_3$ – BaZrSe$_3$, with the direct band gap varying from 1.9 – 1.5 eV, all within the orthorhombic corner-sharing structure type.[16] These prior results were achieved by direct synthesis from the metals Ba and Zr in the presence of H$_2$S and H$_2$Se gases.

Here, we present an alternative method to make BaZr(S,Se)$_3$ chalcogenide perovskite alloys, by annealing BaZrS$_3$ in H$_2$Se gas. This anion-exchange mechanism leaves the perovskite crystal structure intact, and (somewhat surprisingly) has little-to-no effect on the microstructure. This mechanism yields alloy thin films with lower concentrations of extended defects and one-hundred times (100 × ) stronger photoconductive response compared to alloy thin films that we make by direct synthesis.

Anion exchange mechanisms have been developed for many semiconductor alloys, including binary chalcogenides and halide perovskites.[17–21] The use of $H_2S$ and $H_2Se$ gases to tune chalcogenide alloy composition has been developed at an industrial scale for $Cu(In_xGa_{1-x})(S,Se)_2$ (CIGS) thin-film photovoltaics.[22] $BaZrS_3$ is by far the most-studied chalcogenide perovskite, and its synthesis science is advancing rapidly.[1,23–27] Tuning the band gap of $BaZrS_3$ will be essential to develop chalcogenide perovskite photovoltaics. We are hopeful that the results reported here may be combined with ongoing advances in $BaZrS_3$ synthesis to accelerate the development of $BaZr(S,Se)_3$ alloy thin film device technology, leveraging industry know-how in selenization.

## 2. Methods

We used a chalcogenide molecular beam epitaxy (MBE) system (Mantis Deposition M500) to grow and anneal films. Substrates were heated radiatively by a SiC filament and were rotated at 2 rpm; all temperature reported here are according to the heater thermocouple and are systematically higher than the substrate temperature. Ba metal was supplied from an effusion cell (Mantis Comcell 16–500), and Zr metal from an electron-beam evaporator (Telemark 578). We calibrated Ba and Zr source rates using a quartz crystal monitor at the substrate position. We further calibrated source and growth rates after film processing using X-ray reflectivity (XRR), X-ray photoelectron spectroscopy (XPS), and X-ray fluorescence (XRF). Sulfur and selenium were supplied in the form of $H_2S$ and $H_2Se$ gases (Matheson) at 99.9% and 99.998% purity (at the cylinders), respectively, and were further purified by point-of-use purifiers (Matheson Purifilter) before entering the MBE chamber. The gases were injected in close proximity to the substrate using custom-made gas lines and nozzles. We controlled gas delivery using mass flow controllers (Brooks GF100C). We measured reflection high-energy electron diffraction (RHEED) data using a 20 keV, differentially-pumped electron gun (Staib) and a digital acquisition system (k-Space Associates, kSA 400).

We deposited films on $10 \times 10 \times 0.5$ mm$^3$ $(001)_{PC}$-oriented $LaAlO_3$ substrates (MTI), and on 2" diameter (001)-oriented $Al_2O_3$ substrates (MTI); here, PC stands for pseudo-cubic, where $(001)_{PC}$ corresponds to the (012) family of reflections in a rhombohedral system. We began by outgassing the substrates in the MBE chamber at 1000 °C in $H_2S$ gas, before growing a layer of $BaZrS_3$, approximately 40 nm thick, using methods as reported previously.[9]

After $BaZrS_3$ film growth, select samples were removed from the chamber and exposed to air during characterization, before returning to the chamber for selenization. Of these, the film on 2" $Al_2O_3$ was annealed in 0.35 sccm $H_2Se$ at 800 °C for 20 minutes, and the film on $LaAlO_3$ was annealed under a mixture of 0.2 sccm $H_2S$ and 0.2 sccm $H_2Se$ at 800 °C for 60 minutes.

Other samples remained in the chamber for selenization without intervening air exposure. In these cases, we annealed in $H_2Se$ immediately after $BaZrS_3$ deposition without adjusting the substrate heater. We processed three samples in this way, with varying selenization conditions: 0.6 sccm $H_2S$ and 0.1 $H_2Se$ for 37 minutes; 0.3 sccm $H_2S$ and 0.36 $H_2Se$ for 30 minutes; and 0.5 $H_2Se$ for 20 minutes. All three samples were annealed at 1000 °C. These particular $H_2S/H_2Se$ ratios were as used in our previous study of direct growth of $BaZr(S,Se)_3$ alloys.[16] We determined the selenization times by watching for an unchanging RHEED pattern, consistent with perovskite film growth; at longer times, the RHEED pattern was observed to degrade. We maintained gas flows during cooldown after growth to avoid S and Se desorption.

We photographed the 2" wafer samples using a Nikon DSLR at ISO 100, with a 50 mm lens, an aperture of f/2, and a shutter speed of 1/100$^{th}$ of a second. We surrounded the samples by white light diffuser boards, with a lighting color temperature of 3500 K. We photographed a color correction card under the same conditions, to adjust the raw photograph files with the proper white balance and color adjustments.

We carried out X-ray diffraction measurements on the 2" $Al_2O_3$ samples using a Bruker D8 Discover general-area detector diffraction system (GADDS) with a Co $K_\alpha$ source, ¼ Eulerian cradle, and Vantec-2000 area detector. We collected data from 2θ = 20° - 80, at 0°, 30°, and 60° tilt angles. These data were integrated to generate one-dimensional 2θ scans. The background was measured on a substrate without a film and was subtracted from the data. We adjusted the d-spacing data for small variations in sample height offset using $Al_2O_3$ peaks as reference. To make these corrections, we first converted the data from 2θ to d-spacing, made adjustments, and then converted back to 2θ.

We carried out X-ray diffraction measurements on the $10 \times 10 \times 0.5$ mm$^3$ $LaAlO_3$ samples using a Bruker D8 high-resolution X-ray diffractometer with a Ge (022) four-bounce monochromator in parallel-beam mode, with a Cu $K_\alpha$ source at a tube power of 1.6 kW (40 kV, 40 mA). Vegard's law estimates were performed using a reference $BaZrS_3$ literature pattern (ICSD #23288, $d_{(202)} = 2.4899$ Å) and an epitaxial sample ($y = 2.21 \pm 0.06$, $d_{(202)} = 2.642 \pm 0.016$ Å) characterized by scanning transmission electron microscopy (STEM) energy dispersive X-ray spectroscopy (EDS). We determine film composition by considering the ratio between Ba, Zr, S, and Se signals measured by EDS We estimate the errors in measured composition ($y$) by adding in quadrature the errors in determining XRD peak positions, and errors in the reference EDS composition measurements.

We performed spectroscopic ellipsometry (SE) on the 2" wafer samples using a J.A. Woollam M-2000D ellipsometer with rotating compensator and XLS-100 head source unit. Measurements were performed on mirror-smooth surfaces at an angle-of-incidence of 70° in the photon energy range 1.24 to 6.20 eV (1000 to 200 nm).

We used Semilab Spectroscopic Ellipsometry Analyzer (SEA) software to perform model-based analysis of SE data. The optical model consists of the $Al_2O_3$ substrate layer and the $BaZr(S,Se)_3$ film layer. The bulk optical values of these layers are used as initial conditions for performing best-fit regression to the data. The $Al_2O_3$ substrate uses a Cauchy model with Urbach tail for the optical properties. The substrate oscillator parameters were obtained by fitting SE data measured on a bare $Al_2O_3$ substrate. $BaZr(S,Se)_3$ is modeled with 10 Tauc-Lorentz oscillators as outlined by Nishigaki *et al.*[4] (Further modeling details can be found in the **Supplemental Information**)

We prepared cross-sectional samples for STEM measurements by nonaqueous mechanical polishing (Allied PurpleLube). Final polishing was performed by Ar+ ion milling (Fischione 1051 TEM Mill). STEM imaging was performed using a probe aberration-corrected Thermo Fisher Scientific Themis Z S/TEM at an accelerating voltage of 200 kV, semi-convergence angle of 18 mrad, and probe current of 30 pA. Calculation of the spacings between atom columns in STEM images was carried out using a custom Python-based Gaussian fitting routine. STEM EDS was acquired at a probe current of 150 pA and collected using a Thermo Fisher Scientific Super-X detector. The Thermo Fisher Scientific Velox software package was used for EDS quantification.

We carried out photoconductivity spectroscopy (PCS) measurements on samples prepared by sputtering 5 nm Ti/200 nm Au interdigitated contacts on the films. The interdigitated contacts had eight individual fingers, each 5 mm long, with finger spacing and width of 100 μm. We used a 300

W Xe arc lamp tunable light source (ScienceTech), providing irradiance between 50 to 200 µW/cm$^2$ (measured with a thermal power sensor, ThorLabs S310C) at the sample plane, depending on the selected wavelength. Photocurrent spectra were measured using a lock-in amplifier (Stanford Research Systems SR830) to source voltage and measure peak-to-peak photocurrent under mechanically chopped 7 Hz illumination. Our previously-reported PCS measurements used a higher-bandwidth, quasi-DC technique with a slow illumination on/off sequence.[16] To enable quantitative comparison to these previous measurements, we also used a source-meter (Keithley 2400C) to measure the quasi-DC responsivity to 600 nm illumination.

## 3. Results

Selenization by anion exchange works for large-area films and on both epitaxial and non-epitaxial, polycrystalline thin films. In **Fig. 1** we demonstrate synthesis of a BaZrS$_3$ polycrystalline film on a 1-side polished 2" Al$_2$O$_3$ wafer, and selenization after air exposure. The selenized film is much darker than the as-grown sulfide, consistent with a reduced band gap. In **Fig. 1b** we present the results of SE measurements, recorded at the center and edge of the wafer, measured on the same sample before and after selenization. The data are consistent with a reduction in $E_g$ of 0.4 eV (see below). In **Fig. 1c** we present XRD measurements recorded at three positions across the film, measured on the same sample before and after selenization. After selenization, all film peaks have shifted to lower values of 2θ, consistent with a larger $d$-spacing due to selenium substituting for sulfur. The peak shifts are nearly consistent across the wafer, and there is no evidence of a change in crystal texture, grain size, or phase. Using Vegard's law, we estimate that the selenized film in **Fig. 1** has composition BaZrS$_{3-y}$Se$_y$, $y = 1.50 \pm 0.67$.[16]

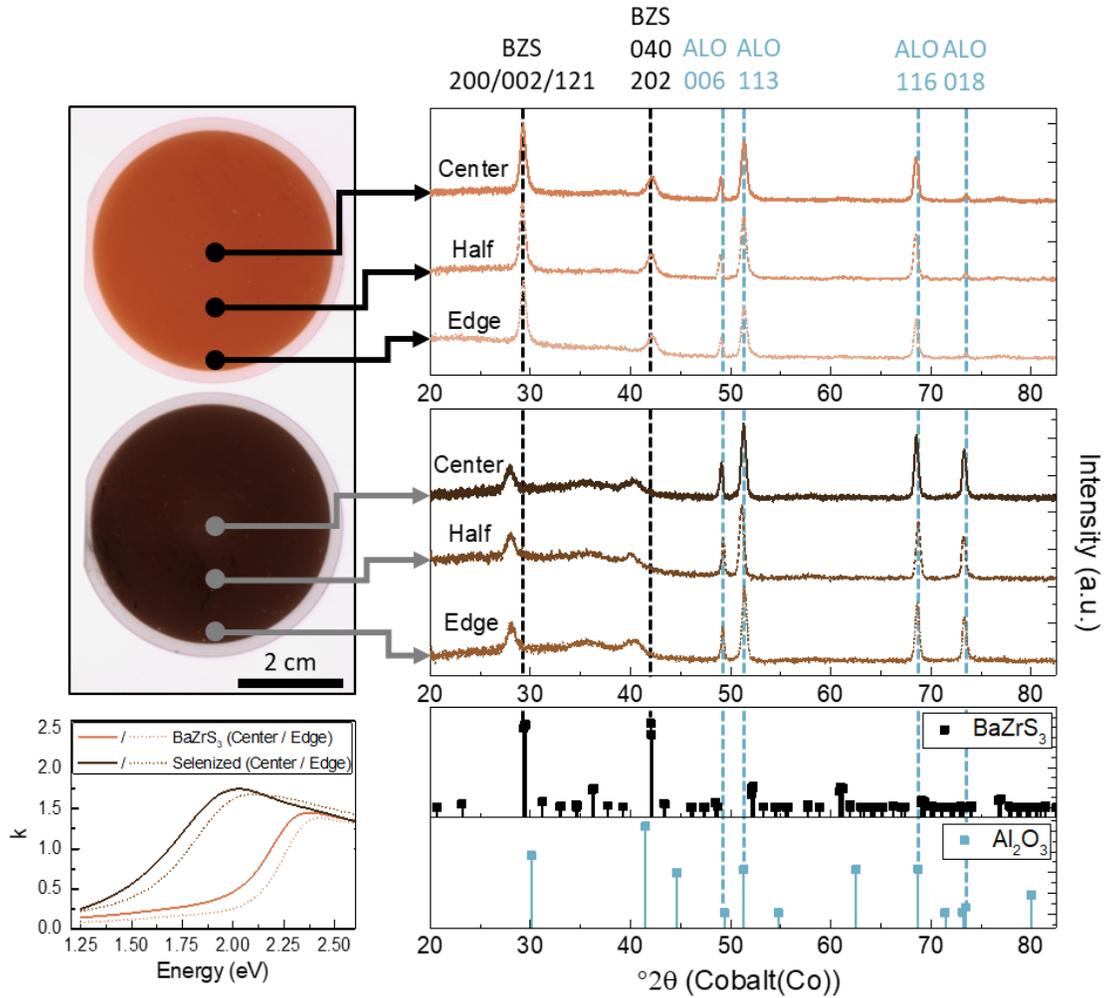

**Figure 1:** Pre- and post-selenization characterization of a polycrystalline BaZrS$_3$ (BZS) thin film. (a) BaZrS$_3$ thin film samples grown on 2" Al$_2$O$_3$ (ALO) wafers before (top) and after (bottom) selenization. (b) Extinction coefficient ($k$) measured by SE on the same sample before and after selenization, at the wafer center and edge. (c) XRD measurements on the same sample before and after selenization, at various points on the wafer. Indexing to BaZrS$_3$ (ICSD #23288) and Al$_2$O$_3$ (ICSD #30030) provided for reference.

The alloy film composition and band gap can be tuned by adjusting the gases used during selenization. In **Fig. 2** we present results of selenizing epitaxial BaZrS$_3$ films grown on (001)$_{PC}$ LaAlO$_3$ with varying H$_2$S and H$_2$Se gas flows. XRD data show that the (202) reflection of BaZrS$_3$ shifts to lower 2θ with increasing H$_2$Se and decreasing H$_2$S flow rates, as expected for selenium substituting for sulfur. Using Vegard's law, we estimate $y = 1.07 \pm 0.39$, $2.05 \pm 0.40$, and $2.24 \pm 0.56$ for the three samples presented. The film also darkens with increasing Se content, although the effect is less apparent in photographs than for the 2" wafer (**Fig. 1**) due to the smaller sample size, different substrate back side polishing, and different photographing conditions. RHEED data recorded during selenization, and atomic force microscopy (AFM) measurements performed after selenization, both indicate that the selenization leaves intact the smooth growth surface, with a root-mean-square roughness of approximately 1 nm for all samples (**Supplemental Information**).

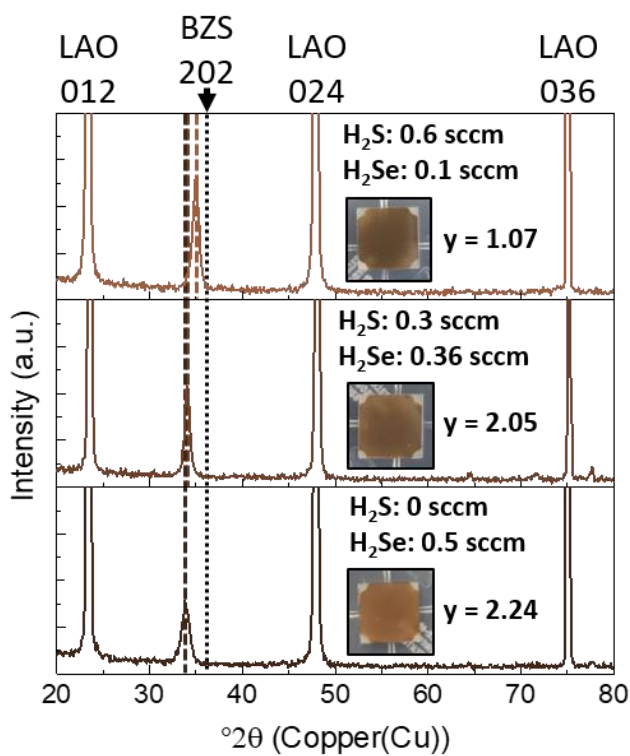

**Figure 2:** Controlling the extent of selenization of epitaxial BaZrS$_3$ (BZS) films grown on LaAlO$_3$ (LAO) by annealing in different H$_2$S/H$_2$Se gas mixtures. XRD results show an increase in $d$-spacing of the BZS (202) reflection with increasing selenization. The composition $y$ of the resulting BaZrS$_{3-y}$Se$_y$ films is determined using Vegard's law. The black dotted line indicates the position of the BaZrS$_3$ (202) reflection; the colored dashed lines indicate the shifted peak positions for the three films shown here. (Insets) Photographs show each of the three samples after selenization.

In our previous result of direct, epitaxial growth of BaZr(S,Se)$_3$ alloys, we observed doublets of the (202) reflection due to a high concentration of antiphase boundaries (APBs).[16] These extended defects – which can also be modeled as isolated layers of a Ruddlesden-Popper phase coherent with the surrounding perovskite lattice – proliferated due to time-dependent film stress during growth. However, these doublets are not observed in the XRD data in **Fig. 2**, despite having similar compositions and the same film-substrate epitaxial relationship as in the previous work. To understand this difference in film microstructure, we turn to STEM.

In **Fig. 3a**, we present high-angle annular dark field (HAADF) data, measured along the $[010]_{\text{LAO-PC}}$ zone axis an epitaxial film of $BaZrS_{3-y}Se_y$ ($y = 2.21 \pm 0.06$) that was selenized after exposure to atmosphere. The film structure is fully-relaxed (*i.e.*, unstrained by epitaxy), as we reported previously for epitaxial growth of $BaZrS_3$ on $LaAlO_3$.[9] As expected from the XRD data, we observe a large reduction in APB concentration, relative to alloys made by direct growth. For comparison, we reproduce in **Fig. 3b** HAADF data measured on a film of comparable composition ($y = 2$) made by direct growth, making apparent the significant reduction in APBs in the selenized film.[16] The distance between adjacent Ba columns in the out-of-plane, [101] direction is $5.286 \pm 0.128$ Å, consistent with the out-of-plane separation of $5.284 \pm 0.032$ Å measured by XRD. We characterize the composition of the film shown in **Fig. 3a** with STEM EDS, and show the results in **Figs. 3c-d**. The EDS maps (**Fig. 3c**) indicate a homogeneous mix of Ba, Zr, S, and Se. We observe a native oxide layer at the top surface, several nm thick, visible in the HAADF and O EDS data. In **Fig. 3d** we present the depth dependence of the atomic concentrations. The Se concentration is uniform through the film and there no detectable segregation of S/Se. The EDS data show a small, uniform concentration of O in the film. This is most likely the result of oxidation of the exposed cross-section during STEM sample preparation and/or secondary florescence.

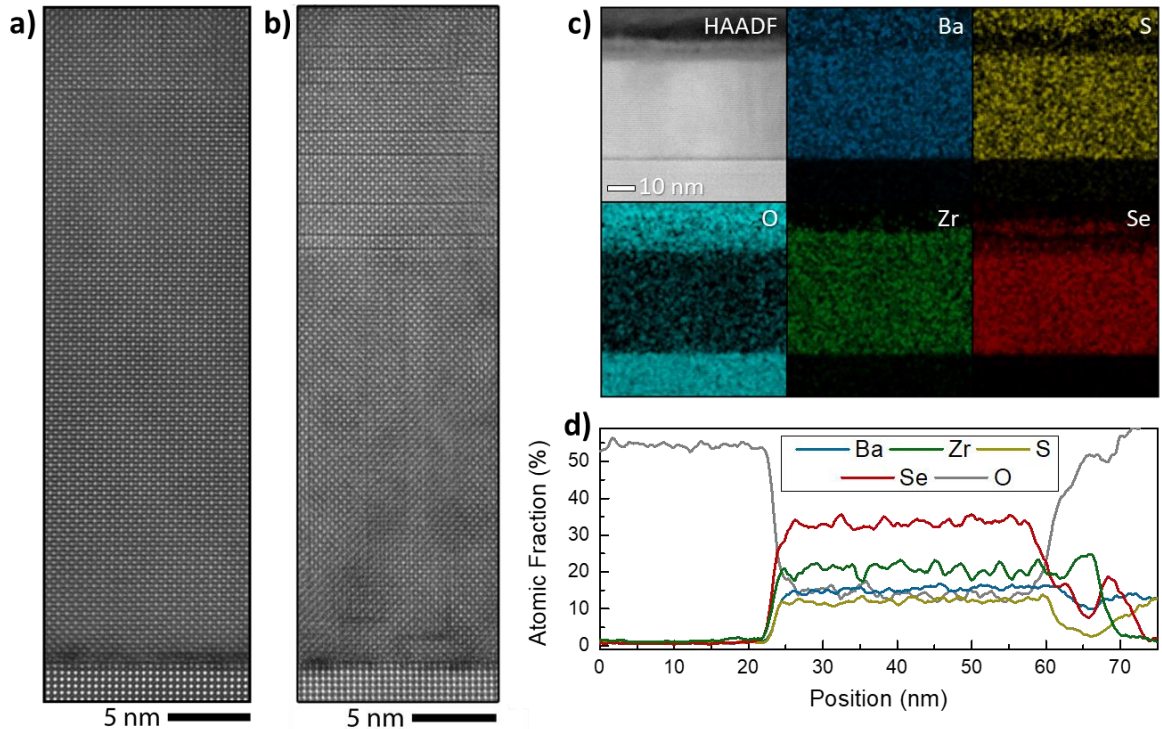

**Figure 3:** HAADF STEM data for an epitaxial film of $BaZrS_{3-y}Se_y$ ($y = 2.21$) grown on $LaAlO_3$. (a) Atomic-resolution image of a typical film cross section showing a noticeably reduced density of antiphase boundaries compared to a direct-growth film (b) HAADF image of direct-growth film ($y = 2$), reproduced.[16] (c) EDS maps of elements Ba, S, O, Zr,

and Se correspond to the HAADF STEM image in (a). (d) Elemental atomic fraction, as determined by EDS, through the film thickness.

In **Fig. 4**, we show results of PCS measurements of an epitaxial BaZrS$_{3-y}$Se$_y$ ($y = 2.21$) film selenized after exposure to atmosphere and the same sample seen on STEM in **Fig. 3**. In **Fig. 4a**, we plot the normalized responsivity to highlight the shift in absorption edge upon selenization, from $1.9 \pm 0.1$ eV for BaZrS$_3$ to $1.5 \pm 0.1$ eV for the selenized sample. We plot in **Fig. 4b** the absolute responsivity of the same selenized film. In **Fig. 4c**, we show that the results here for the trend of band gap *vs*. composition are consistent with our previous results of direct synthesis. In **Fig. 4d** we compare the responsivity of the alloy film made by post-growth selenization to results of direct synthesis. The responsivity of the sample made in this study is approximately 100 × higher than our previous results. The responsivity depends on the film optical density, and on the lifetime ($\tau$) and mobility ($\mu$) of optically-excited charge carriers. The $y \approx 2$ films represented in **Fig. 4d** are comparable in composition and thickness, and identical in phase, and therefore the 100 × change in responsivity cannot be due to a difference in optical density and must be ascribed to increases in $\tau$ and/or $\mu$. We hypothesize that this improvement in excited-state charge transport is due to the reduced density of APBs, which may act as recombination and/or charge scattering sites.

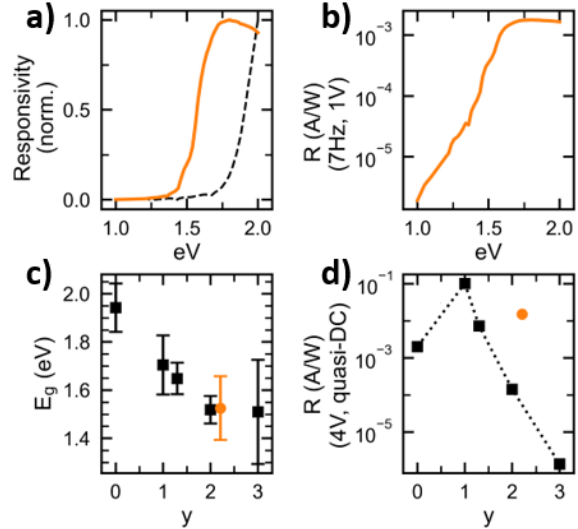

**Figure 4:** PCS measurements of epitaxial BaZrS$_{3-y}$Se$_y$, ($y = 2.21$) film made by selenization. (a) Normalized responsivity of the selenized sample, showing a band gap reduction of 0.4 eV compared to the pure sulfide. (b) Responsivity ($R$) spectrum measured using a lock-in method at 7 Hz. In (c-d), the black data are previous results of direct alloy film growth, and the orange data point is the sample analyzed here.[16] (c) Trend of band gap, estimated from PCS data, *vs*. Se-content. (d) Responsivity at 600 nm illumination, measured with a slow illumination on/off sequence at 0.3 Hz. We attribute the 100 × increase in responsivity of the selenized alloy film, compared to the film made by direct growth, to the substantial reduction in extended defect concentration.

## 5. Discussion and Conclusion

Our results demonstrate the effectiveness of anion-exchange processing using an $H_2Se$ atmosphere on both polycrystalline and epitaxial $BaZrS_3$ thin films to create $BaZr(S,Se)_3$ chalcogenide perovskite alloys. We find that post-growth processing can control the alloy composition and band gap of these materials without disturbing the crystal structure. This results in samples that have a significantly lower density of extended defects compared to samples made by direct growth, and we attribute this finding to our improved photoconductive response.

Theory predicts that the orthorhombic, corner-sharing perovskite structure is thermodynamically unstable at high Se content, and that $BaZrS_{3-y}Se_y$ will instead adopt face-sharing or edge-sharing structures with reduced band gap well below 1 eV.[28,29] Our previous work demonstrated that $BaZrS_{3-y}Se_y$ epitaxial thin films are stable in the perovskite structure in the full composition range, but left unaddressed the question of whether non-epitaxial films would likewise be stable perovskites. The results here demonstrate that polycrystalline thin films, without epitaxy, can also be stable perovskites at with high Se content and reduced band gap. Earlier results found that direct reaction of Ba, Zr, and Se powders results in the formation of a non-perovskite, vacancy-ordered, hexagonal phase.[30,31] The stability of the films reported here with respect to such competing phases requires further study. In our experience, we find that these films are stable for prolonged periods of storage in ambient conditions, as evidenced for instance by indistinguishable XRD data recorded eight months apart.

$BaZrS_3$ is air-stable, but upon air exposure and storage does form a thin native oxide, that we have observed to be several nm thick in our MBE-grown films.[9] The presence of this oxide layer can be seen in RHEED: the RHEED pattern of the epitaxial perovskite film, which remains distinct throughout film growth, disappears when the film is handled in air for some time and re-introduced into the MBE chamber. However, despite this interruption to the crystal structure, we find that selenization proceeds readily with or without air exposure between film growth and annealing in $H_2Se$. We hypothesize that the native oxide is unstable and quickly degrades at the annealing conditions. Whatever the exact cause, this is encouraging for further studies of selenization of $BaZrS_3$ samples made by different means, and it bodes well for future large-scale processing of perovskite chalcogenides in solar cell technology.

**Acknowledgements**

We acknowledge support from the National Science Foundation (NSF) under grant nos. DMR-1751736 and DMR-2224948. This research was supported in part by a grant from the United States-Israel Binational Science Foundation (BSF), Jerusalem, Israel. This research was supported in part by the Sagol Weizmann-MIT Bridge Program. This research was supported in part by the Skolkovo Institute of Science and Technology as part of the MIT-Skoltech Next Generation Program. K.Y. and J.V.S. acknowledge support by the NSF Graduate Research Fellowship, grant no. 1745302. M.X. and J.M.L acknowledge support from the MIT Research Support Committee. This work was carried out in part through the use of the MIT Materials Research Laboratory and MIT.nano facilities. We acknowledge Sunil Mair for assistance with sample photography.

**Supplementary information for "*A Processing Route to Chalcogenide Perovskites Alloys with Tunable Band Gap via Anion Exchange*"**

Kevin Ye[1*], Ida Sadeghi[1*], Michael Xu[1], Jack Van Sambeek[1], Tao Cai[1], Jessica Dong[1], Rishabh Kothari,[1] James M. LeBeau[1], R. Jaramillo[1‡]

1. Department of Materials Science and Engineering, Massachusetts Institute of Technology, Cambridge, MA 02139, USA

*These authors contributed equally to this work

‡ *email address:* rjaramil@mit.edu

**S1: General-area detector diffraction system (GADDS) XRD post-processing**

In **Fig. S1**, we show the raw data and detail the post-processing of the general-area detector diffraction system (GADDS) data. We found that our polycrystalline samples do not produce full arcs in the GADDS scans, as would be expected for sample with completely randomly oriented grains. Instead, we notice preferred orientations (*i.e.*, texture) of BaZrS$_3$ *Pnma* (220)/(022), (240)/(042), (123)/(321), and (440)/(044) aligned with the (001) reflections of Al$_2$O$_3$. The BaZrS$_3$ reflections are listed in pairs because they are indistinguishable at the resolution of this diffractometer.

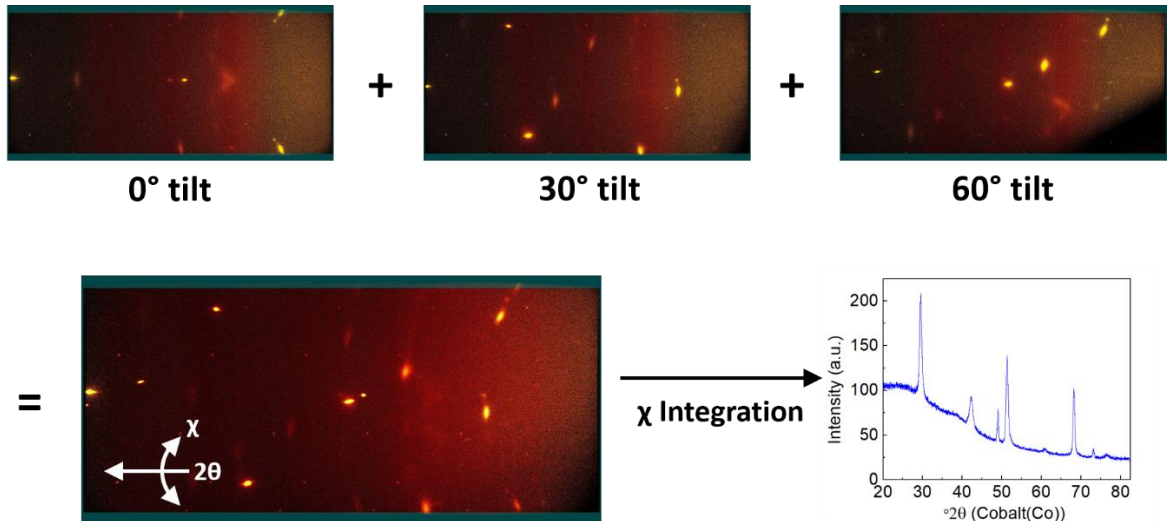

**Figure S1:** Post-processing of the raw data from the general-area detector diffraction system (GADDS), for measurements taken at the center of a polycrystalline BaZrS$_3$ film on a 2" Al$_2$O$_3$ wafer. Three different scans were taken at tilt angles of 0°, 30°, and 60°. These frames were overlaid and then integrated at each value of χ to generate one-dimensional 2θ scans.

The film texture increases the difficulty of reliable phase identification using $\theta - 2\theta$ XRD scans with a point detector and highlights the utility of the GADDS measurements. We collected a wide range of spots across multiple frames, which are then synthesized into 2θ scans. In **Fig. S2**, we show the background obtained from the methodology shown in **Fig. S1** and subtracted from the 1D 2θ scans.

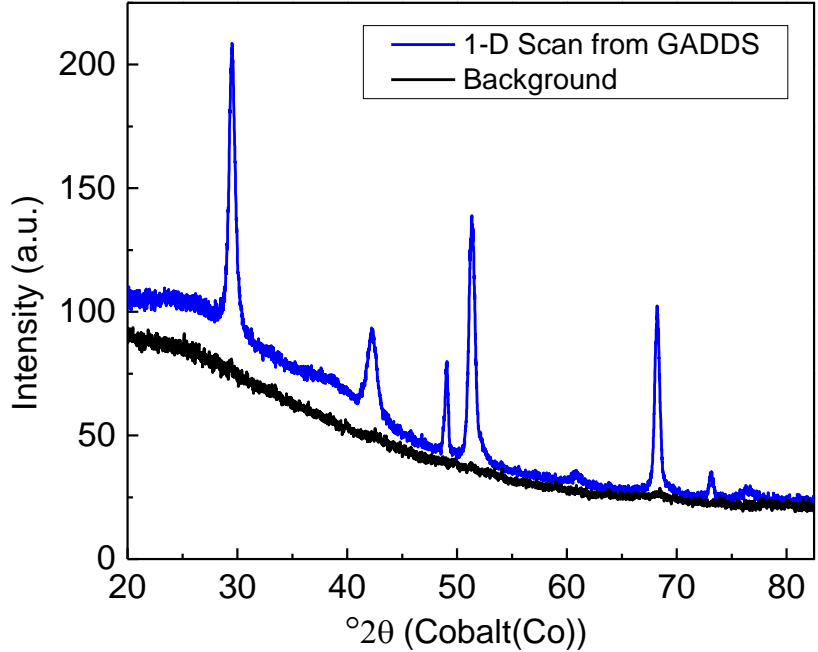

**Figure S2:** Background subtraction of the 2θ data series synthesized from GADDS data. The background data was obtained from a scan on substrate without a film and subtracted from the data.

## S2: Spectroscopic ellipsometry modeling

We found that the $Al_2O_3$ substrate undergoes slight changes in optical properties following film growth due to exposure to high temperatures and reactive $H_2S$ and $H_2Se$ gases. This observation is further supported by XRD, which suggests that the $Al_2O_3$ lattice distorts after film growth. To accommodate this, the we perform data regression in two sequential steps: first, the substrate alone undergoes one round of fitting using the Levenberg-Marquardt algorithm (LMA), and then the film is separately fit starting again using LMA. The SEA software enables this multi-step isolated fitting process, and our analysis found that this procedure produces the highest fidelity fits to the data. The Tauc-Lorentz parameters for our $BaZrS_{3-y}Se_y$, ($y = 1.85$) sample determined by this method are in **Table S1** below:

| Oscillator | A (eV) | $E_0$ (eV) | C (eV) | $E_g$ (eV) |
|---|---|---|---|---|
| 1 | 216.59959 | 4.30186 | 0.81826 | 4.42423 |
| 2 | 15.7537 | 2.20243 | 1.72333 | 0.62953 |
| 3 | 18.54995 | 2.87353 | 1.24114 | 2.36588 |
| 4 | 5.26347 | 4.41684 | 0.80671 | 2.92375 |
| 5 | 3.56928 | 3.97564 | 1.16381 | 2.73655 |
| 6 | 0.67523 | 3.54061 | 0.43585 | 2.58892 |
| 7 | 8.01904 | 0.55599 | 0.23812 | 2.73373 |
| 8 | 0.9384 | 2.08165 | 0.28213 | 1.09893 |
| 9 | 26.6922 | 1.83603 | 0.35093 | 1.33601 |
| 10 | 6.53535 | 8.52778 | 0.12729 | 2.71366 |

**Table S1:** Spectroscopic ellipsometry (SE) fit parameters for 10 Tauc-Lorentz oscillators to model the $BaZrS_{3-y}Se_y$, ($y = 1.85$) sample, grown on $Al_2O_3$.

## S3: X-ray diffraction of epitaxial film characterized by STEM and photoconductivity spectroscopy

We present in **Fig. S3a** high-resolution XRD (HRXRD) data of an epitaxial $BaZrS_{3-y}Se_y$ film (y = 2.21), selenized after exposure to atmosphere. These data illustrate that removal of the sample from the MBE, and likely formation of a thin native oxide, does not interfere with subsequent selenization. The out-of-plane scans for the films show the (202) reflection, shifted to lower $2\theta$ relative to $BaZrS_3$, indicating a larger $d$-spacing as Se substitutes for S in the crystal structure. In **Fig. S3b**, we show a phi scan which shows the in-plane symmetry of the sample, further showing that the selenization treatment retains the crystal structure. We find that the film (121), (200), and (002) peaks are aligned with the substrate (110) peaks, without any additional peaks. This in-plane epitaxial relationship is the same as we reported previously for $BaZrS_3$ on $LaAlO_3$. For this epitaxial sample, we measure the film in-plane peaks with the detector position at $2\theta = 24.598°$.

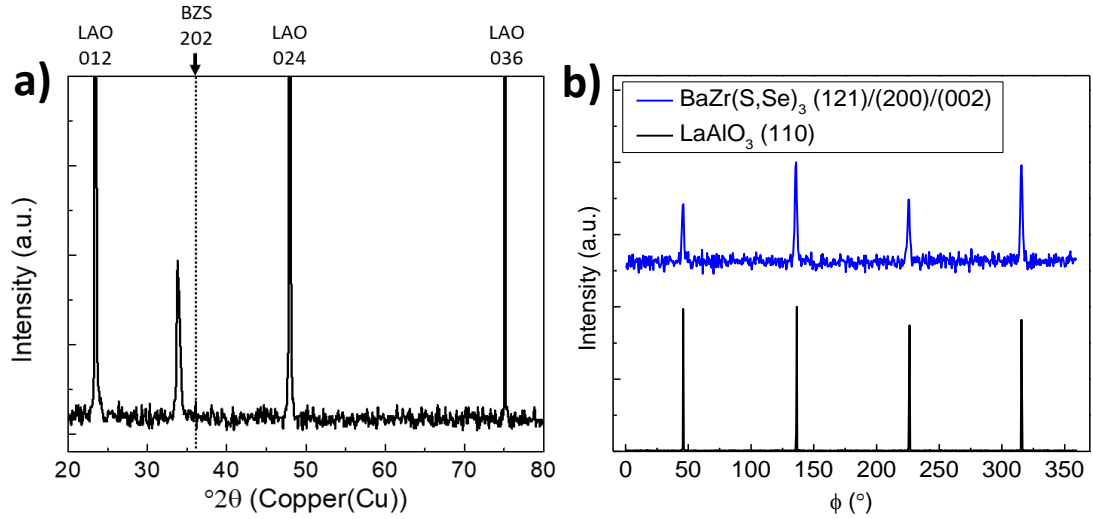

**Figure S3:** HRXRD of epitaxial BaZrS$_{(3-y)}$Se$_y$ ($y = 2.21$) alloy grown on LaAlO$_3$ (LAO). (a) Out-of-plane scan with the position of the (202) reflection of the pure sulfide BaZrS$_3$ (BZS) marked for reference. (b) Phi scan showing the in plane epitaxial relationship between BaZrS$_{(3-y)}$Se$_y$ and LaAlO$_3$. The film (121)/(200)/(002) peaks are aligned with the substrate (110) peaks.

### S4: Film surface characterization

We present in **Fig. S4** reflection high energy electron diffraction (RHEED) and atomic force microscopy (AFM) on epitaxial BaZrS$_{(3-y)}$Se$_y$ alloy samples post-selenization. The presence of RHEED streaks indicate an atomically-smooth, crystalline surface. The AFM data root-mean-square roughness ($R_q$) for each alloy is approximately 1 nm.

We present in **Fig. S5** characterization of the surface of our polycrystalline samples using atomic force microscopy (AFM) and scanning electron microscopy (SEM). Although the root-mean-squared roughness of the increases by 2 nm with our selenization treatment, we find that the SEM suggests a smoother surface for the selenized sample when looking at a wider field of view. We attribute this observation to the higher, but approximately constant Z height of the grains observed in **Fig. S5b** when compared to those in **Fig. S5a**.

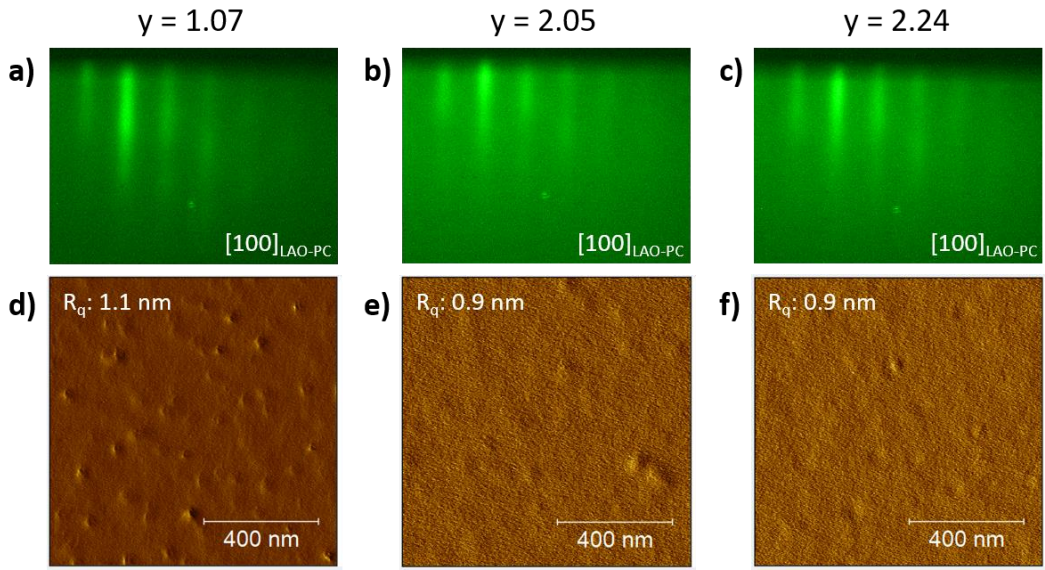

**Figure S4:** RHEED (a-c) and AFM (d-f) of epitaxial BaZrS$_{(3-y)}$Se$_y$ ($y$ = 1.07, y = 2.05, y = 2.24) alloys grown on LaAlO$_3$ (LAO).

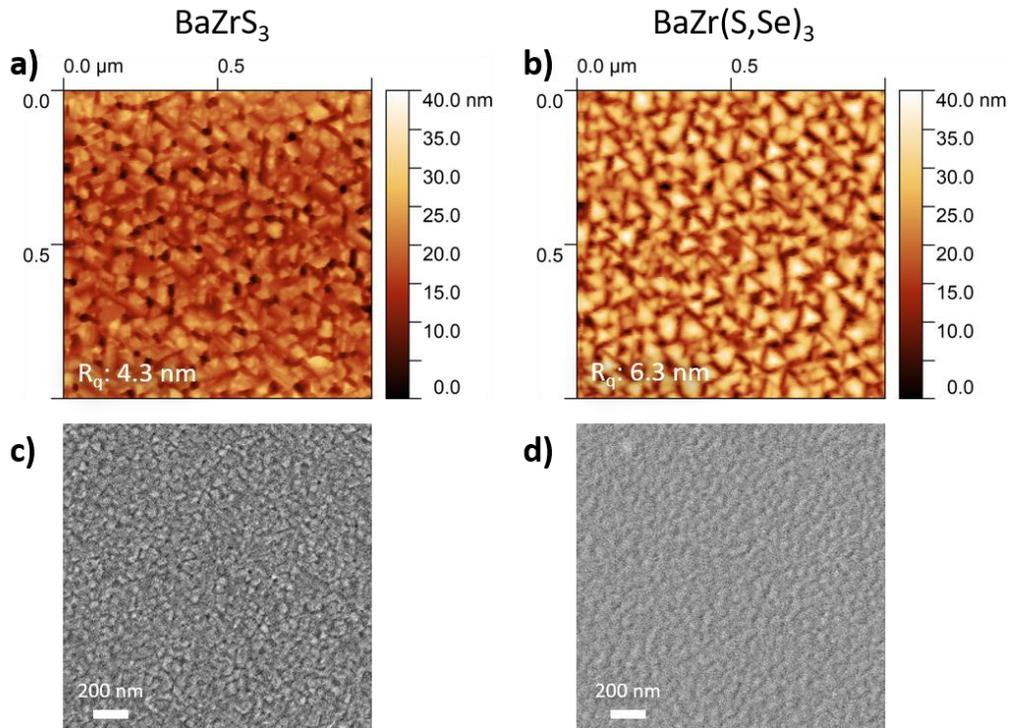

**Figure S5:** Atomic force microscopy (AFM) (a-b) and scanning electron microscopy (SEM) (c-d) of polycrystalline thin film samples grown on 2" sapphire wafers. A pristine BaZrS$_3$ sample is shown (a,c) for comparison with a selenized sample (b,d).

## S5: STEM-EDS Sample Composition

In **Table S2** we present the film atomic fractions determined from STEM EDS for an epitaxial sample. The approximate composition is $BaZrS_{3-y}Se_y$ ($y = 2.21$). We then use this measured composition to estimate the composition of other films, combining with XRD data and using Vegard's law. Due to the potential for uncontrolled surface oxidation during STEM sample preparation and experimental artifacts in quantifying O, we exclude O in the calculation.

| Z | Element | Family | Atomic Fraction (%) | Atomic Error (%) |
|---|---|---|---|---|
| 8 | O | K | 14.82 | 0.90 |
| 16 | S | K | 12.05 | 2.08 |
| 34 | Se | K | 35.01 | 3.27 |
| 40 | Zr | K | 22.51 | 2.53 |
| 56 | Ba | K | 15.62 | 1.70 |

**Table S2:** STEM EDS composition results for elements observed measured through the film thickness.

## S6: Responsivity Band Gap Estimation

We present in **Fig. S6** the band gap estimation methodology used in the photoconductivity spectroscopy analysis. This analysis employs a linear fit of the responsivity data near the absorption onset, rescaled in the manner of direct band gap Tauc analysis. The analysis assumes responsivity is from the photogenerated carriers only, meaning that the responsivity is directly proportional to absorption coefficient. Estimated band gap of $1.5 \pm 0.1$ eV is given by the x-intercept of the linear regression, and the uncertainty represents a 95% confidence interval calculated from regression standard errors.

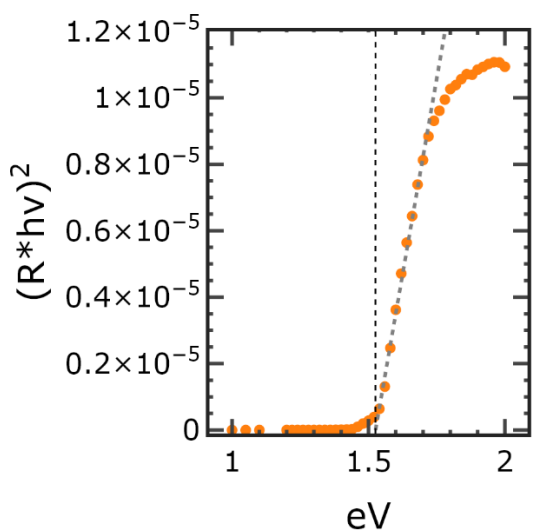

**Figure S6:** Tauc-style plot of data (orange circles) and fit (dashed line) for the epitaxial BaZrS$_{3-y}$Se$_y$ ($y$ = 2.21) film's responsivity spectrum. The estimated band gap is 1.5 ± 0.1 eV.

### S7: Table of samples reported here

In **Table S3** we list the samples reported in this work, and their names as appearing in our laboratory notebooks and database. The initial BaZrS$_3$ growth step for all samples was performed under 0.8 sccm H$_2$S at 1000 °C.

| Sample description | Substrate | Database name | Growth/Annealing Conditions |
|---|---|---|---|
| Polycrystalline BaZrS$_3$ | 2" Al$_2$O$_3$ | G144 | BaZrS$_3$ growth for 45 minutes |
| Polycrystalline BaZrS$_{(3-y)}$Se$_y$ ($y$ = 1.50 ± 0.67) | 2" Al$_2$O$_3$ | G107 | BaZrS$_3$ growth for 81 minutes<br>Annealing Conditions:<br>• 0.35 sccm H$_2$Se<br>• 800 °C<br>• 20 minutes |
| Epitaxial BaZrS$_{(3-y)}$Se$_y$ ($y$ = 2.21 ± 0.06) | 10*10 mm LaAlO$_3$ | G101 | BaZrS$_3$ growth for 60 minutes<br>Annealing Conditions:<br>• 0.2 sccm H$_2$S/0.2 sccm H$_2$Se<br>• 800 °C<br>• 60 minutes |
| *Samples annealed under H$_2$S/H$_2$Se atmosphere without prior exposure to atmosphere* | | | |
| Epitaxial BaZrS$_{(3-y)}$Se$_y$ ($y$ = 1.07 ± 0.39) | 10*10 mm LaAlO$_3$ | G123 | BaZrS$_3$ growth for 60 minutes<br>Annealing Conditions:<br>• 0.6 sccm H$_2$S/0.1 sccm H$_2$Se<br>• 1000 °C<br>• 37 minutes |
| Epitaxial BaZrS$_{(3-y)}$Se$_y$ ($y$ = 2.05 ± 0.40) | 10*10 mm LaAlO$_3$ | G120 | BaZrS$_3$ growth for 60 minutes<br>Annealing Conditions:<br>• 0.3 sccm H$_2$S/0.36 sccm H$_2$Se<br>• 1000 °C<br>• 30 minutes |
| Epitaxial BaZrS$_{(3-y)}$Se$_y$ ($y$ = 2.24 ± 0.56) | 10*10 mm LaAlO$_3$ | G122 | BaZrS$_3$ growth for 60 minutes<br>Annealing Conditions:<br>• 0 sccm H$_2$S/0.5 sccm H$_2$Se<br>• 1000 °C<br>• 20 minutes |

**Table S3:** Samples reported in this work and processing conditions